# A Brief History of the Statistics Department of the University of California at Berkeley.

by Terry Speed, Jim Pitman and John Rice.

The early history of our department was dominated by Jerzy Neyman (1894-1981), while the next phase was largely in the hands of Neyman's students, with Erich Lehmann (1917-2009) being a central, long-lived and much-loved member of this group. We are very fortunate in having the biography *Neyman – From Life* written by Constance Reid (1918-2010), and Erich's *Reminiscences of a Statistician: The Company I Kept* and other historical material[1] documenting the founding and growth of the department, and the people in it. In what follows we will draw heavily on these sources, describing what seems to us a remarkable success story: one person starting[2] "a cell of statistical research and teaching…not being hampered by any existing traditions and routines," and seeing that cell grow rapidly into a major force in academic statistics worldwide. That it has remained so for (at least) the half-century after its founding is a testament to the strength of Neyman's model for a department of statistics.

## 1. The years before the department (1938-1955) and the department.

In 1938, at the age of 44, Jerzy Neyman left the Department of Applied Statistics at University College, London and arrived in Berkeley to take up a position as Professor of Mathematics. His status as a leading scholar of his generation was firmly established by his paradigm-setting work in areas including hypothesis testing, confidence intervals, and sampling design. Shortly after Neyman's arrival, he founded the Statistical Laboratory (Stat Lab), and this was to be the center of Statistics at Berkeley until the creation of a separate Department of Statistics in 1955, and the center of his activities until his death in 1981.

As well as throwing himself into theoretical and applied statistical research, Neyman devoted a great deal of energy to teaching – for example, during the second semester of 1939-40 he increased his voluntary teaching load to 25 hours per week. In 1939 he recruited Elizabeth (Betty) Scott (1917-1988, BA 1939), an astronomer who had been auditing his course, as a research assistant, while in 1941 Evelyn Fix (1904-1965), a high-school mathematics teacher who had attended a summer session at Berkeley, joined the Lab as a technical assistant. Statistical work contributing to the US effort in World War 2 soon became the principal focus of the Lab, in particular extensive research on bombing patterns.

When the war ended in 1945, Neyman was still the only Berkeley faculty member in the field of statistics. Well aware of the war efforts of many outstanding mathematicians and mathematical statisticians elsewhere in the US, he organized a symposium on mathematical statistics and probability[3] "to contribute to the revival of scientific work in mathematical statistics and allied fields". This landmark gathering featured contributions from Hotelling, Pólya, Wolfowitz, Doob, Hsu, Feller, and Neyman's student, Erich Lehmann, as well as papers on the application of statistics to physics, astronomy, psychology, economics, forestry, animal breeding, rainfall, and insect populations. Hotelling's keynote address "The Place of Statistics in the University[4]," and the ensuing discussion are as relevant today as they were then.

The *Berkeley Symposium on Mathematical Statistics and Probability* was a great success, and was to be the first of six, held roughly every five years, whose proceedings grew from a single volume of 500 pages published in 1949 to six volumes totaling well over 3,000 pages published in 1972. Plans began almost immediately for a second symposium, to take place in 1950. These symposia were an[5] "enormous achievement for Neyman, a testament to his imagination, energy and organizing ability. During the twenty-five years of their existence, they functioned as the most important international statistics meeting, and put Berkeley on the map as a world center."

Neyman returned with enthusiasm to teaching in Fall 1945, assembling a varied group of individuals to cover ten statistics courses. He led the team, and was assisted by Pao-Lu Hsu (1910-1970), his former student and later colleague from University College, London. At that time, they were the only two instructors with PhDs, and they were joined by Erich Lehmann, by then a graduate student in mathematics, who received his PhD under Neyman in 1946, as did Evelyn Fix (PhD 1948), John Gurland (1917-1997, PhD 1948) and Mark Eudey (1912-2007, PhD 1949). Two others came from outside statistics to complete the team: Elizabeth Scott, who received her astronomy PhD in 1949, and Edward Barankin (1920-1995), who completed a mathematics PhD on linear algebra in 1946.

Four of the Fall 1945 instructors went on to become faculty in Mathematics and later Statistics, Lehmann in 1946, Barankin in 1947, and Fix and Scott in 1950, and another Neyman student from this era, Joseph L. Hodges Jr (1922-2000, PhD 1949) in 1951. Neyman persuaded the campus administration that he had good reasons to go against their policy of discouraging departments from hiring their own PhD graduates, arguing that few other universities were turning out students he considered to be satisfactory. Lucien Le Cam (1924-2000, PhD 1952) was the last of Neyman's students to be appointed (in 1952) to the faculty. These six Berkeley graduates were joined by four outstanding outside appointments. Charles Stein (Columbia PhD 1947), a student of Abraham Wald, was appointed in 1947, Michel Loève (1907-1979), a student of Paul Lévy came in 1948 from a position at the University of London and a visiting appointment at Columbia, Henry Scheffé (1907-1977) whose PhD was in mathematics, but had moved into Statistics, and spent the academic year 1946-7 in Berkeley as a Guggenheim fellow, returned permanently in 1953, and David Blackwell (1919-2010), a student of Joseph Doob, visited Berkeley in 1954-55, and joined the department in 1955. We think of these ten as the department's "founders", as all but one was present when the department was founded in 1955, and remained there until their retirement or death.

We will discuss teaching and research in Statistics within the Stat Lab and later Department in separate sections below. One notable event prior to formation of the Department was the 1949 California oath controversy, where faculty were required to sign an anti-communist loyalty oath. As is recounted in the books by Reid and Lehmann, this had a profound impact on the Stat Lab, and resulted in the permanent loss of Charles Stein to the University of Chicago and later Stanford. The Berkeley symposia continued as planned, the second in 1950 and the third in 1954-55, and Neyman even started his own journal, *University of California Publications in Statistics,* modelled on the *Statistical Research Memoirs* which he had earlier founded with E. S. Pearson at University College London. The *Publications in Statistics* ran from 1949 until 1963, and contained the PhD theses of many Berkeley statistics graduates from that period, and several important papers by Stat Lab faculty, including pioneering papers by Le Cam, but very few papers by outsiders.

The story of Neyman's relentless push for the creation of a Department of Statistics separate from Mathematics has been told before[6], and we will not recount it again here. Suffice to say that Griffith C Evans (1887-1973), the long-running chair of Mathematics at Berkeley who had brought Neyman to Berkeley in the first place[7], and had supported him in most of his efforts after his arrival, had an opposing vision, one of a broad-based Department of Mathematics encompassing all of the mathematical sciences.

On Evans' retirement in 1949, the way became clear for the formation of a separate department, and it eventually came into being in the fiscal year 1955-56.

*The Department*
Less than a year into the life of the new department, its Chair, Jerzy Neyman offered his resignation, to be effective July 1, 1956. Again this is a story that has been told well elsewhere[8], and again we pass over the details, though the following gives a hint as to the reasons[9]

"his [Neyman's] tenure [as chair] was short and bumpy. There were conflicts with the new dean…over a variety of issues including teaching loads and the role of statistical consulting. In addition a breach had opened up between Neyman and some of his oldest colleagues over some internal issues."

When the inevitability of Neyman's resignation became apparent, the unanimous view was that David Blackwell should become Chair, and so he did, to the complete satisfaction of everyone involved. Neyman remained as Director of the Stat Lab, an operation with a research budget at the time about equal to the department's budget, and would remain so until his death.

Space at a university is a stand-in for power and has long been a flashpoint for campus unhappiness and disputes. As soon as Neyman set up the Stat Lab, infrastructure was required: technical assistants, computing equipment, offices and space. Neyman obtained all that. When the department was formed, more space was required. Eric Lehmann has described the Department back then as being "constantly on the move". The graduate students typically had office space in a collection of so-called T-buildings, T for temporary. These were left over from the World War II years, and lasted until the 1990s. However in the sixties discussions began with the NSF about the funding of a "mathematics" building.

In 1971 Evans Hall (named after the long-running chair of Mathematics) opened. Betty Scott was one of the people much involved in its planning. Initially Statistics had the third and fourth floors in the building. There was a group of offices for the Stat Lab, with Neyman having a suite for his scientific papers, his secretary, meetings and for small classes. The Department administration also had a group of offices. The graduate students were now all housed together with the faculty in Evans. On the third floor there was a lounge that could double as a seminar room, six classrooms, a library, a reprint room, and a pair of rooms for an electronic computer and its peripheral equipment. Eventually the campus administration took back half of the fourth floor. They had started out by seeking to take a complete floor.

*Biostatistics.*
The Interdepartmental Group Biostatistics was created in 1955 by Neyman and Jacob Yerushalmy (1904-1976), linking the biostatisticians in Berkeley's School of Public Health (SPH) with statisticians in the Department of Statistics for the purpose of granting PhD degrees in Biostatistics. For a period the Group had co-chairs, one from the SPH and one from Statistics. The initial ones were Chin Long Chiang who completed his Statistics PhD under Neyman in 1954, and Betty Scott. Today the biostatisticians are located within the Division of Epidemiology in the SPH, and the Group continues to intertwine the SPH and the Department, both students and faculty.

*Social Life.*
Since Evans Hall opened a lounge on the tenth floor has served as both a Statisitics class/seminar room and as a gathering place. It was christened the Neyman Room in the early 1980s and contains some mementos of Neyman's life and career. The lounge on the third floor serves as a lunch room, as a coffee room and as a gathering place. There are regular picnics and sometimes intramural teams in soccer and baseball. These may be organized by the Statistics Students Graduate Association (SGSA) or by a faculty member. There is an annual t-shirt design competition, sometimes won by a student and sometimes by a faculty member. There are also get-togethers of ex-students and faculty at professional meetings such as the JSM.

*Visitors*

There have been many visitors to the department over the years. In particular, Berkeley's Miller Institute for Basic Research in Science has invited many statisticians as Visiting Professors, including Leopold Schmetterer (1919-2004) in 1959, Herbert Robbins (1915-2001) in 1966-67 and Frederick Mosteller (1916-2006) in 1974-75. Berkeley also has the prestigious Hitchcock Lectures, given approximately annually by a distinguished visitor. Since their inception, three statisticians have presented these lectures and visited the department: Ronald A Fisher (1890-1962) in 1936, when it was still the Department of Mathematics, John W Tukey (1915-2000) in 1975, and Frederick Mosteller in 1985. Several visitors came (and continue to come) regularly, including Florence Nightingale David (1909-1993). David had known Neyman in the 1930s at University College, London, where she received her doctorate in 1938. After a career at UCL, culminating in being appointed Professor in 1962, David joined the University of California at Riverside, and taught there until she retired in 1977. She had begun making regular summer visits to Berkeley in 1948, where she taught in summer session, and these visits continued well after her retirement.

In the next three sections, we will summarize the department's efforts at teaching, research, and consulting, calculating and computing, from this early period up until the present day. Then we will describe where we are now, and where we are going.

## 2. Statistics teaching at Berkeley

In his first year at Berkeley, Neyman taught two classes, one on upper division probability and statistics that had been regularly offered in the past, and a second on new developments, including the work of von Mises, Kolmorogov and the Neyman-Pearson theory. This may have been a little ambitious, but as time went on, Neyman's ambitions did not lessen. In the Introduction to his 1950 book *First Course in Probability and Statistics* "for a one-semester basic course for beginners" he wrote "As conceived in this book, the theory of statistics is a section of the theory of probability." He went on to acknowledge his "indebtedness to his friend and colleague, Professor Erich L. Lehmann, who conceived the idea that the basic concepts of the modern statistical theory can and should be taught at an elementary level. Also, he was the first to try this idea in class, with considerable success." It is of some interest to note that Neyman's *First Course* began with an introductory chapter concerning the scope of the theory of probability and statistics which referred to inductive behavior, statistical decision functions, admissible hypotheses and performance characteristics, and which was followed by a chapter introducing probability in a more or less traditional manner, with permutations, combinations, conditional probabilities and independence. However, this second chapter ended with a quite serious 25-page section entitled *Evaluation of Competing Risks*. Chapter 3 was entirely devoted to *Probabilistic Problems of Genetics,* and began with an explanation of the laws of heredity, including recombination. It was clear from the outset that for Neyman, the teaching of statistics at Berkeley was not to be a rehearsal of cookbook-style recipes for the stylized analysis of data. However, *the best laid schemes o' mice an' men…*.

Neyman's book was written for students taking Stat 1, which for many years was the "theoretical" introductory course, in contrast to the "non-theoretical" introductory course Stat 2. In his *Reminiscences* Lehmann recounts Scheffé's memorable explanation of the distinction[10]: "the more theoretical course (Stat

1) was for students who needed to understand statistics but were not planning to use it, while the methods oriented cookbook-style course (Stat 2) was for students who planned to use statistics but did not need to understand it."

Undergraduate statistics texts at Berkeley, of which Neyman's was the first, show how introductory statistics teaching there has evolved. Joe Hodges was a superb and inspiring teacher who was particularly concerned with undergraduate teaching. In this connection he co-authored the elementary texts *Basic Concepts of Probability and Statistics* (1964, 2nd ed. 1970) jointly with Lehmann, which was translated into Danish, Hebrew, Italian and (fairly recently) Farsi, and *Stat Lab - An Empirical Approach* (1975) jointly with the psychologists Krech and Crutchfield. Teaching became difficult when he suffered severe hearing loss and he retired in 1991.

Hodges and Lehmann's *Basic Concepts* was a successor to Neyman's book, arguably one more accessible to the Stat 1 audience[11]. Blackwell's *Basic Statistics* (1970) was for Stat 2, while *StatLab* was another introduction, to a different audience again. By that time Stat 1 had been split into separate courses in probability and statistics, but Stat 2 has always remained. The most recent and by far the most successful of all these introductions is *Statistics* by David Freedman (1938-2008), Robert Pisani and Roger Purves, first edition 1978, and now in its fourth (2007) edition. This unique text is neither "methods-oriented" nor "cookbook-style," but while making minimal mathematical demands on its readers, it[12] "teaches a different approach: thinking." Many of us believe it succeeds admirably at doing so. For more than 30 years now, the book has been used by over 2,000 students at Berkeley each semester, in Stat 2 and other introductory courses.

Hodges and Lehmann's *Elements of Finite Probability* (1965) was the first to separate probability from statistics, and this was followed by Aram Thomasian's *Structure of probability theory with applications* (1969), which was for electrical engineers, and *Probability* (1993) by Jim Pitman.

The teaching of large lower-division courses such as Stat 2 has rarely been a popular assignment among regular faculty, notable exceptions including Blackwell, Freedman and Jack Kiefer (1924-1981). As a result, the department has appointed lecturers, that is, faculty whose principal duty is teaching, to fill this role. Roger Purves and Ani Adhikari are our current senior lecturers and Hank Ibser our continued lecturer. Together with some regular faculty, visitors and occasionally senior graduate students, they organize large groups of teaching assistants, and ensure that our service teaching is at a very high standard. Many of the ladder faculty are superb teachers as well, notable examples being Blackwell, Freedman, and Deborah Nolan. Adhikari and Nolan have both received the extraordinarily competitive Distinguished Teaching Award, given each year to between 1 and 5 members of Berkeley's ~1,500 full-time faculty. In 2007 Philip Stark presented the first on-line course at Berkeley. He continues to develop his on-line statistics course, called SticiGui, and adapt it to different introductory audiences.

*The major*
The department has offered an undergraduate major since 1955-56, and for most of that time, majors in a given year numbered in the teens. Over the last decade, enrolments have exploded -- there are currently about 250 majors (many of them double majors, particularly with Mathematics and Economics). Nolan has been instrumental in building our statistics major. She has published several books on teaching statistics and in the mid-2000s pioneered a very popular upper-division course called Concepts in Computing with Data. In addition to this course, and standard courses on such topics as linear models, stochastic processes and time series analysis, there are small interactive seminar courses on special topics which emphasize student projects and presentations. With the impetus of an NSF VIGRE program, many undergraduates participate in faculty research, often working alongside graduate students and postdocs.

### *The quals*
Over the years, the form of the doctoral qualifying exam (quals) has evolved. In the early 1960s the first year curriculum for Ph.D. students was uniform: students enrolled in *two* year-long courses (Probability Theory and Theoretical Statistics), and took oral exams in *three* areas: Probability, Theoretical Statistics and Applied Statistics. Preparation for the last-mentioned was the reading of Scheffé's *The Analysis of Variance*. In the early 1980s a year-long PhD-level course in Applied Statistics was introduced. Incoming PhD students now enrolled in two of the three year-long courses, and took written quals in their two at the end of the summer following their first year. Students who failed the quals exams twice left the program.

Since about the year 2000, it became increasingly apparent that this one-size-fits-all curriculum was a poor match to the increasing diversity of faculty and student interests and of preparation of students entering the program. For example, there was general agreement that some students would be better served by being able to take a course in, for example, Mathematics, Computer Science, or Biology during their first year and defer some core Statistics courses until their second year. Also we found that only a very small number of students who had left the program did so because of failing the quals twice (students did leave, but for other reasons). In 2005 the uniform curriculum and written quals of the first phase of the program were replaced by a more flexible system with evaluations each semester being based on course work and coordinated by the Head Graduate Advisor (see http://www.stat.berkeley.edu/programs/phd for details). Rather than study for quals during the first summer, students engage in reading and research. The new system appears to be working well. We can expect pressure further change around 2025-2030.

### *Text books*
Over the years Berkeley faculty have produced some widely-used and influential graduate-level probability and statistics texts. Loève's 1955 book on probability theory and Lehmann's 1959 and 1983 books on testing and estimation, respectively, are arguably the best known and most widely-used ones. (A mimeographed preliminary version of Lehmann's estimation book had been circulating for almost 30 years prior to the publication of the actual book.) A full list of books by the founders and selected books by later generation faculty is appended below.

## 3. Research at Berkeley

### *Probability*
A student of the distinguished Russian probabilist Serge Bernstein, Neyman had studied Lebesgue's theory of measure and integration before Kolmogorov built the measure-theoretic foundations of modern probability theory in 1937 on an abstract form of Lebesgue's theory. Neyman's interests shifted quickly from measure theory to the foundations of statistics, but he continued while at Berkeley to work with Scott on probability models for the clustering of point processes, motivated by applications in astronomy. Neyman also maintained an appreciation of the mathematical theory of probability, which was developed rapidly during the period when he influenced appointments at Berkeley. He attracted to Berkeley representatives of the two main schools of probability theory in the early 20th century: Blackwell, a student of Doob, and Loève, a student of Paul Lévy. Both Blackwell and Loève held joint appointments in Statistics and Mathematics, and probability thrived at Berkeley at the interface of these two departments.

Further appointments built up the probability group at Berkeley to include Lester Dubins (1920-2010), David Freedman and P. Warwick Millar. Blackwell, Dubins and Freedman collaborated during the 1960s to produce a series of influential results around inequalities for martingales and related stochastic processes, while Millar developed these ideas to provide foundations for the theory of stochastic integration for processes with independent increments.

The tradition of probability at Berkeley became widely known especially through the influential text books on probability and stochastic processes by Loève, his student Leo Breiman (1928-2005) and by Freedman. By the 1970s Berkeley was established as one of the best known departments internationally for research and exposition of probability. This reputation attracted a next generation of probabilists to Berkeley in the late 1970s and early 1980s, when Michael Klass, Jim Pitman and David Aldous joined the department. During the 1980s, Klass obtained definitive results around the law of the iterated logarithm. Aldous's work on mixing times of finite Markov chains provided inspiration for a renaissance of the theory of finite Markov chains and its applications to computer science and computational statistics through MCMC. Pitman collaborated extensively with Marc Yor at Paris VI on the theory of Brownian motion and related stochastic processes. In the late 1980s and early 1990s Aldous developed his general theory of continuum random trees as limits of combinatorial models of random trees. While on sabbatical in Paris in 1989, Pitman worked with Jacques Neveu and Jean-Francois Le Gall to develop a theory of random trees embedded in Brownian excursions. It was some years before the confluence of these ideas was fully appreciated through collaboration of Aldous and Pitman on asymptotic theory of random trees and associated tree-valued and partition-valued random processes. Throughout the 1990s and 2000s there was an ongoing exchange of ideas between the Berkeley and Paris schools. Motivated by applications to Brownian motion and Bessel processes, Pitman and Yor developed enrichments of the theory of random discrete distributions (Dirichlet distributions), with roots going back to work of Blackwell and Ferguson in the early 1970s. Le Gall in Paris applied random trees to the construction of Markovian superprocesses, a theme developed further in Berkeley by Steven Evans (1989). Aldous and Pitman worked on the theory of continuum random trees, which motivated further formalizations of the theory of random real trees by Le Gall and Evans. Aldous, Evans and Pitman worked on various aspects of processes of coagulation and fragmentation processes and their relations to random trees, much of which theory which was put in a definitive form by Jean Bertoin. Yuval Peres joined the department in the late 1992. His research across a wide range of modern probability theory, including particle systems and random walks, Brownian motion, percolation and critical phenomena, attracted a large number of students and injected much energy into the Berkeley probability group in the early 2000s. More recent appointments in probability at Berkeley include Elchanan Mossel (2003) and Allan Sly (2010), with interests at the interface of probability and computer science, and Sourav Chatterjee (2006, Stein's method, theory of spin glasses, random matrices).

In 1992 the *Line and Michel Loève International Prize in Probability* was created in honor of Michel Loève, by his widow Line. Awarded every two years, the Loève prize is intended to recognize outstanding contributions by researchers in mathematical probability who are under 45 years of age. This biennial award is a notable recurrent event involving the Berkeley probability group.

**Statistical theory at Berkeley, the founders.**
Theoretical statistics at Berkeley in the decade from the mid 1940s was dominated by the contributions of Erich Lehmann, Joe Hodges and Lucien LeCam, together with Henry Scheffé and Charles Stein, though Neyman himself still had unfinished business. Underlying themes included the continuation of Neyman's agenda to find tests that were optimal in multi-parameter settings, the development of testing and later estimation within Wald's emerging framework of statistical decision theory, and an increasing focus on large-sample properties of tests and estimators, arising in part from a searching critique of the method of maximum likelihood.

Neyman's main contribution was to develop the theory of Best Asymptotically Normal estimates, for both estimation and testing, with a major paper being presented at the 1945 Berkeley Symposium, prepared in part as early as 1941. Later, he devised the class of $C(\alpha)$ tests for testing in the presence of nuisance parameters, a topic he was to continue working on until the end of his life. Interestingly, this class of tests has recently (2011) reappeared providing an approach to uncovering rare genetic risk factors from DNA sequence data.

Lehmann's collaboration with Scheffé began while the latter was a Guggenheim fellow in 1946-7, and in the following years produced their now classic work on completeness, similar regions and unbiased estimation. Lehmann also discovered the notion of complete class of decision rules in the course of proving results concerning admissibility of tests, and later discussed general unbiasedness. In 1948 Neyman and Scott provided some compelling examples that reinforced an observation made earlier by Wald on the inconsistency of the maximum likelihood estimator in a context where the number of nuisance parameters grows along with the number of observations. This was followed not long afterwards by Hodges' 1951 discovery of a superefficient estimator of the mean of a normal distribution with known variance. This counterexample, together with seminal work of Wald, paved the way for Le Cam's PhD thesis. There he proved that for one-dimensional parameters, Bayes estimates exhibit local asymptotic minimaxity and admissibility, and showed that the corresponding maximum likelihood estimates inherit both properties by being asymptotically Bayes. In addition, he discussed superefficiency further, and proved that the superefficiency points form a Lebesgue null set in the parameter space. Soon afterwards (1955), Le Cam published his extension of Wald's theory of statistical decision functions, and went on to define the Locally Asymptotically Normal condition, the notions of contiguity, distance (deficiency) between experiments, approximate sufficiency, limits of experiments and much more. With the hindsight afforded by a perspective of over 50 years, there seems little doubt that among the most lasting achievements of the Berkeley school in this period are those of Le Cam, who was, in the words[13] of Grace Yang (PhD 1966), "the principal architect of the modern asymptotic theory of statistics."

Other noteworthy contributions of the founders from among many include Hodges and Fix's pioneering 1951 report on nonparametric density estimation, Scheffé's 1953 discovery of what became known as the S-method of multiple comparisons, Lehmann and Stein's deep 1948 paper "Most powerful tests of composite hypotheses", the Hodges and Lehmann body of research on nonparametric estimators and tests, and Barankin's many papers on the theory of sufficient statistics and exponential families.

*Statistical theory: non-founders.*
The next generation of statistical theorists began with Peter Bickel (PhD 1963) and Kjell Doksum (PhD 1965), both students of Erich Lehmann, who joined the faculty as assistant professors in 1963 and 1966 respectively. Both have broad interests, and they collaborated on the analysis of transformations, revisiting the famous work of Box and Cox on this topic, and in their well-known graduate level text *Mathematical Statistics*. Bickel wrote on robust estimation and was a participant in the Princeton Robustness year (1970-71). He also made extensive contributions to distribution free methods, minimax, Bayes and later adaptive estimation, semi-parametric models, bootstrap asymptotics, higher-order expansions, and much more. More recently, topics he has worked on include hidden Markov models, regularization methods and particle filters. Doksum's best known work was on tail-free and neutral random probabilities, which later linked up with work of Berkeley probabilists, and he also contributed to reliability theory, survival analysis and local correlation.

The year 1970 saw the appointment of Rudolf Beran, who worked in many of the areas listed above for Bickel, but who continued working on different aspects of the bootstrap and adaptive estimation for much longer, collaborating with Millar on minimum distance estimation and several other topics. In 1977 Ching-Shui Cheng started in the department as an assistant professor. He had written his Cornell PhD thesis on optimal experimental design under Jack Kiefer, the person who almost single-handedly created that field, and in 1979 Kiefer himself joined the department. Kiefer's was a very strong theorist with broad interests, and he was also a highly successful instructor in lower division courses. His time in Berkeley was sadly cut short by a sudden, fatal heart attack in August 1981. Cheng has remained the department's sole person in experimental design, contributing extensively to several different aspects of this topic.

In 1980 Leo Breiman was appointed, see next section, and Charles Stone shortly afterwards. Their joint work on classification and regression trees (CART) and Stone's on nonparametric function estimation and splines took the department into new directions, and this movement continued with the appointment of David Donoho in 1984. Major interests of Donoho at that time were sparse modelling, signal recovery and inverse problems, particularly in geophysics, and he worked on these topics together with a postdoctoral fellow, Philip Stark, who was to join the faculty in 1988, and continue in these areas long after Donoho left Berkeley for Stanford.

Later appointments in the area of statistical theory include Deborah Nolan (1987), who originally worked on U-processes and Bin Yu (1993), whose Berkeley thesis was on information theory and empirical processes. Both of them have moved away from pure theory, in Nolan's case towards statistics teaching, including writing books and developing new courses in this area, and working towards gender equity in the mathematical sciences. Yu remains interested in theory related to the applied problems she attacks, and her current theoretical interests include sparse modelling and machine learning. These last interests are shared with Bickel and Michael Jordan (appointed 1998), Peter Bartlett (2002) and Martin Wainwright (2004), the last three joint being with Electrical Engineering/Computer Science (EE/CS). Our most recent appointments in statistical theory are Noureddine El Karoui (2005), with interests in random matrices and mathematical finance, and Aditya Guntuboyina (2012), who works on applications of convexity to statistics, minimax lower bounds, and a variety of other topics.

For many years Berkeley was generally unsympathetic to the Bayesian approach to statistics, with the notable exception of Blackwell. Freedman had long been interested in the consistency and inconsistency of Bayes estimators, but in the last few years, Michael Jordan and Cari Kaufman (appointed 2008) have made more positive contributions to Bayes theory and methods.

*Information theory*
In the 1950s and early 1960s Blackwell, Breiman and Thomasian made several key contributions to information theory, perhaps the most notable being Breiman's 1957 proof of the individual ergodic theorem of information theory. More recently Yu and Wainwright have both worked on different aspects of coding theory.

**Applied Statistics at Berkeley**

Jerzy Neyman came to Berkeley in 1938 with considerable experience applying statistics and a very strong commitment to linking theory and applications closely. An English translation from the original Polish of his 1923 PhD thesis title was *Justification of applications of calculus of probability to the solution of certain questions of agricultural experimentation.* This set the pattern for all his work. Neyman also regarded applications as an essential part of the statistician's education, commenting at the 1945 Berkeley symposium[14]

"The future mathematical statistician needs early contacts with experimental sciences. He needs them because, at this stage of the development of statistics, the experimental sciences are sources of theoretical problems. Also, he needs them because in almost any imaginable job which he may get after graduation he will be called upon to apply his theory to experimental or observational problems."

We see in his *First Course* how he viewed applications of theorems on probability (pp. 69-70)

"The first difficulty is that no practical problem is concerned with mathematical concepts… Therefore, the practical problem must be translated into probabilistic terms before its solution is attempted…In fact, this translation amounts to building up a mathematical model of the practical problem. …In certain cases, the adequacy of the particular model can be tested empirically…. Naturally, the solution of the problem based on a given model applies to the model itself and not necessarily to the phenomena for which it was intended. ….The second …difficulty …is that most observable phenomena are rather complicated. ….In many cases, …we are forced to revise the model, sacrificing its adequacy in order to gain simplicity."

Neyman reiterated these sentiments many times, and they permeate all his applied work, the sheer volume of which is dizzying. Not long after arriving in Berkeley Neyman made contact with the California Forest and Range Experiment Station located on the campus, and the Department of Entomology. In due course Neyman together with other members of the Stat Lab, especially Scott and Fix, extended their applied research to many other areas, including bacteriology, fisheries, public health, demography, accident proneness, astronomy (over 30 papers), weather modification (over 30 papers), carcinogenesis, molecular evolution and ecology. For some discussion of some of this enormous body of work, and a complete list of papers, we refer to Neyman's obituary[15] and to the published Neyman Lecture, Brillinger (2008) for an overview of three examples.

Of the founders, it was principally Neyman, Scott and Fix who conducted applied research, though in the late 1970s and early 1980s, Le Cam became involved in the cancer research carried out by the Stat Lab, and collaborated with the physicians Byers and Levin in research in cancer immunotherapy. David Brillinger's appointment in 1970 immediately expanded the intellectual interests of the department in ways beyond those represented by the founders, although his work on point processes related to that of Neyman and Scott, and there was a connection to harmonic analysis (Loève). He came with a background in time-series analysis, an interest in point processes, and experience working with his advisor John W.Tukey and others at Bell Labs. Fairly soon after his arrival in Berkeley, Brillinger and his students were collaborating with a neurophysiologist, a seismologist, and with scientists interested in the abundance and movements of wild animals such as whales. Over the years his applied interests broadened to include imaging, spectroscopy, risk assessment, imaging, demography and meteorology.

The next appointee with applied interests was Leo Breiman, who joined the department in 1980. Coming from a second career as a private consultant, following a distinguished academic career at UCLA, it was expected that Breiman would be interested in applications. He was, but unlike Neyman and Brillinger, he was not particularly interested in getting closely involved with scientific collaborations, and he was even less interested in using probability models as the basis for his statistical analyses. Leo's forte was creative ideas for dealing with standard statistical questions, typically with large data sets, methods that are now part of the literature of data mining and machine learning. He was an expert on bootstrapping, cross-validation and other methods used for variable selection in regression, and played a central role in the development of many novel techniques, including classification and regression trees (CART), alternating conditional expectations (ACE), archetype analysis, boosting, bagging, and random forests. These are now very widely used in science, business and industry.

David Freedman joined the department in 1961, ostensibly as a probabilist, but one with a strong and critical interest in statistics, and he remained on the faculty for the rest of his life. His extensive contributions to applied statistics were stimulated by his undergraduate teaching and his consulting. He felt that compelling illustrations of statistical phenomena from the scientific literature would interest students more than "baby" ones, and this was a key aspect of his highly successful undergraduate statistics text. For example, it makes use of the original 1903 Pearson and Lee data relating the heights of fathers and sons in examples, and there is a plot of experimental data (from Berkeley) illustrating Hooke's law. Freedman's 1971 consulting with the Bank of Canada on econometric models began his career as a trenchant critic of much statistical practice in the social sciences, and this was to be a pre-occupation for the rest of his life. For over 35 years, he consulted widely with foundations, commissions, cities, state and federal government departments and agencies, corporations, banks, and law firms, he appeared as an expert witness on many occasions, and gave congressional testimony twice. In most cases he would carry out all the analyses himself, and if not, he would very closely supervise his assistants. He was a critic of the energy modelling of the early 1980s, census adjustment, much econometric modelling, the use of path analysis and other methods claiming to elucidate causal relations in the social sciences, and much more, and his position was invariably reached through critical re-analysis of key data sets. Some indication of the scope of his work can be found in his 2005 book and the 2010 collection of his papers.

Donoho's and Stark's work on inverse problems has been mentioned under statistical theory, while another appointment in this general area was Finbar O'Sullivan, now at University College, Cork, Ireland. Stark's later research includes confidence sets, election auditing, public policy issues and online statistical education. Terry Speed was a senior appointment in 1987, and he began research in statistical genetics and genomics which continues in the department today through the work of Haiyan Huang (appointed 2003), and Elizabeth Purdom (2009), and Sandrine Dudoit in Biostatistics. John Rice was appointed in 1990, further strengthening the department's commitment to applied work. At that time he was collaborating with neurophysiologists, and later moved into interdisciplinary work on transportation and astronomy and into functional data analysis. Bin Yu's interest in the statistical theory of sparse approximation has been complemented by applied work in information technology, remote sensing and neuroscience, while both Yu and Jordan carry out research on text mining. Yun Song and Rasmus Nielsen were both appointed to the department in 2007 via a campus initiative in computational biology, Song jointly with EE, and Nielsen jointly with Integrative Biology. Cari Kaufman shares the interests of Brillinger and Yu in interdisciplinary research in spatial and environmental statistics, as well as working on Bayesian methods in applications.

## 4. Consulting, calculating and computing at Berkeley.

*Statistical Consulting*

Neyman viewed engaging in consulting with faculty in other departments as an essential part of the role of a university statistician. In this regard, he was completely in agreement with Hotelling, who had addressed the 1940 IMS meeting in Hanover, NH that Neyman attended, on the topic "The Teaching of Statistics." Among other things, Hotelling told his audience that[16]

"A specialist in statistics on a university faculty has a threefold function. In addition to the usual duties of teaching and research, there is a need for him to advise his colleagues, and other research workers, regarding the statistical methods appropriate to their various investigations. The advisory function is a highly important one for the activities of the university as a whole, …"

A side comment is worth making here. After hearing Hotelling give this address, Neyman went back to Berkeley determined to throw himself into writing the book which would become his 1950 *First Course in Probability and Statistics*. In addition, he invited Hotelling to speak on the same topic at the Berkeley symposium in 1945. Writing on personal events taking place towards the end of the 1951-52 academic year, Lehmann explains how he saw it then[17],

"An important activity of the [Stat] lab was the consulting service that provided help for faculty members in other departments with their statistical problems. Neyman handled some of these himself; others he passed on to members of the staff. He considered it one of our obligations to accept such assignments."

Thus consulting was part of the Stat Lab experience from the outset, and many statistics grad students during the 1950s and 1960s supported by research grants funded through the Lab found themselves working with Neyman and other faculty in the Lab on campus consulting. It was not until 1977 that things became more formal, after Juliet Popper Shaffer visited the department for the academic year 1976-77, and joined it as a Lecturer in 1977. Shaffer was a psychologist whose research interests centred around mathematical learning theory. However, she had written many papers on the statistical methods used in psychology, such as factor analysis and other multivariate methods, multidimensional contingency tables and multiple comparisons, the last being the topic that had brought her to Berkeley in the first place. But more to the point, Shaffer had extensive experience in statistical consulting, and she was to be fully responsible for the organization of the department's "Statistical Consulting Service" and for the course "Statistical Consulting" for over a decade, before it became shared more widely across faculty with applied interests. The pattern for the service and course has not really changed much since its inception in Fall 1977. Graduate students sign up for it, sometimes multiple times, and they are put in pairs and assigned a weekly 2-hour slot, during which time they are available in the consulting room to all comers. The hours were advertised campus-wide, but nowadays the service occupies a central position on the department's web site. Once a week the class meets as a group with the course instructor, to discuss the problems that have come to them, and these class meetings are frequently preceded by individual discussions between students and the responsible faculty member. There is no obligation on the consultants to do the analyses for the client, but nevertheless, this sometimes happens. It seems to be a good model. Many lasting scientific interactions have come out of these consulting service encounters, as well as the passing on of much good advice on design and analysis of experiments and studies. Initially the campus Graduate Division paid students for their participation, with $6/hour being the rate in 1977, but in the late 1990s, this practice was discontinued.

**Calculating and computing**
Every department like ours has a symbiotic relationship with calculating and computing machinery, and to an extent all histories of this symbiosis will be broadly similar. What differs from place to place will be the nature of the facilities in use at any particular time, the extent of access of faculty and students to these facilities, and the resources devoted to the area, including money for hardware, software and people.

Writing about Evelyn Fix, Lehmann says[17] that "Her first big task came in 1942, when Neyman received a contract for bombing research from the National Defense Research Council. The work was computationally very intensive; it was carried out on desk calculators and consumed much time and effort. Evelyn was in charge of the calculations and presided over a hastily recruited, ragtag group of students (of which I was one for a short while) and faculty wives." In 1945 Neyman felt able to state[19] that in the course of its contract war research, "the Laboratory acquired an efficient set of computing machines and other equipment." Unfortunately we lack first hand information concerning the machines in use at the time, but it seems very likely that in they were electromechanical calculators, either Marchants, from the company founded in 1911 in nearby Oakland, CA, Fridens, made by the former chief designer of Marchant, who by then had his own company, or Monroes. These calculators were also used for in-course labs during the 1950s and 1960s.

In the late 1960s, the eminent industrial statistician Cuthbert Daniel (1904-1997) was visiting Berkeley. He recommended that the department purchase a Wang electronic calculator, and it did, a 320K. It seems likely that Olivettis were also in use at that time. Part of the NSF grant towards the construction of Evans Hall was for a computer to be shared by Mathematics, Statistic and Computer Science, and a committee selected a DEC PDP-11/45, which was installed in 1974. From the beginning the plan was to use the DEC operating system RSTS, but not long before this, a new operating system called Unix had been devised at Bell Labs. Records indicate that RSTS was first booted on our PDP in early June 1974, and Unix a few days later. Thus Berkeley statistics had access to the first or one of the first machines running Unix outside Bell Labs, and it was also to be the first machine outside Princeton running ISP and the first outside Bell Labs running S. At the same time, there was a campus computer center mainframe in the basement of Evans Hall, initially a CDC 6400 and later a succession of Crays, and these were to be there for about 25 years, until the mid-1990s.

By the late 1970s, teletype terminals were installed in the lab rooms to broaden access to the centralized computers. Over the years they were replaced by a succession of terminals with better graphical capabilities, the aim being to have a terminal in every office, one for the exclusive use of each faculty member, and a shared one for grad students. In due course ISP and S were supplemented by GLIM developed by the Royal Statistical Society, GenStat from the Rothamsted Experimental Station in the UK, and the UK-based Numerical Algorithms Group collection of algorithms in FORTRAN.

In 1981 Leo Breiman was appointed Director of the Statistical Laboratory. Soon after he arrived, it was clear to Breiman that[20] "for the department to ever get into any kind of relationship with data, they had to have a decent computing facility." Following that the Department obtained its first VAX, an 11/750, in 1982, with funding provided by a successful ONR proposal prepared by Breiman.

**The Statistical Computing Facility**
In January 1986 the Statistical Computing facility (SCF) was formed with Leo Breiman as the first director. That same year the Department acquired clusters of Sun 3 work-stations and servers, again as a result of Breiman's efforts. The Suns provided, for the first time, large bit-mapped screens with graphical user interfaces. Because they were networked together any user could access any computer. In a fairly short time the SCF staff consisted of the director, a systems manager, who presided over a team of part-time "ops", and with them addressed hardware systems needs of faculty and students, an applications manager/software consultant, and an administrative manager.

The SCF supports both the research and instructional needs of the department. Phil Spector, who came to us from SAS in 1987 as the first applications manager, started a two-course graduate sequence in statistical computing. In the early 1990s a group within the department led by then student D. Mark Abrahams (PhD 1995) developed BLSS (Berkeley Interactive Statistical System), which was used for some undergraduate instruction. Its advantages included that it ran efficiently and had a relatively simple user interface. By this time computing was integrated into most of the courses in the undergraduate major. In the mid-nineties the department also began making extensive use of the statistical software R, developed by one of our ex-students, Ross Ihaka (PhD 1985), and his collaborator, Robert Gentleman. In 2004, Nolan developed an undergraduate computing course required for the major.

The SCF obtained Sun SPARCstations in 1990, Sun Ultra 2 workstations in 1992, and Sun Ultra 5 and 10 workstations in 1999. In 2004 the SCF began migrating from SPARC to x86-64 with the Sun Java workstations. The size and complexity of applications software increased steadily. In 2006, the SCF acquired its first Apple computers with the purchase of 30 iMacs for the undergraduate computing laboratory.

In the late 1980s and 1990s we would continue to hear about the *ad hoc* computing arrangements in other departments, including quite distinguished ones, some of which were almost entirely reliant on shoestring budgets and the kindness of graduate students. Breiman's dynamic leadership and excellent contacts in the industry, support from national funding agencies for hardware, and support from the university for use of computers in teaching have contributed to the successful operation of the SCF. Having thousands of accounts definitely impressed the administration. The growth of personal computing has impacted the SCF, but it remains core to the research and educational mission of the department.

Currently, research in the SCF is distributed over 24 Linux AMD and Intel based workstations, 13 Linux AMD Opteron compute servers, network and communication services, 34 Intel single and dual-core iMac desktop computers located in undergraduate computing laboratories, and 33 other single and dual-core PowerPC and Intel Mac workstations. Statistics students invariably use personal laptops running R.

## 5. Where are we now, and where are we going?

From the perspective of early 2012, the department grew slowly but steadily between the early 1960s and the late 1980s and early1990s. New faculty were hired, our courses continued to evolve, the "sausage machine" continued to produce PhD graduates in large numbers, and we seemed to be adapting to the changing world, which became increasingly driven by computing and an emphasis on applications. The founders aged and began to retire in this period, but the next generation – students of the founders, or their equivalents elsewhere – came into the department and flourished. However, we failed to notice the aging of this generation until it was almost too late. By 2000 the prospect of 10 faculty members retiring in the period 2007-2014 became apparent, a huge change for a modest-sized department. In that year the chair,

John Rice set up a "next decade" committee to examine how the department should respond, over the coming ten years, to the ongoing changes in the discipline of Statistics. This committee began their October 2000 report noting that

"Today's conventional wisdom …is that the future of Statistics consists of interdisciplinary applied research together with some new synthesis of what constitutes core theory" and went on to say that "we envisage new faculty hires as the main engine of change." They recommended that permission be sought from the administration to "borrow" positions against future retirements so that we could hire on average one person per year for the coming decade, and that we introduce the option of "PhD in Statistics with Emphasis in X" for several interdisciplinary subjects X. The report also recommended that the department should place greater value on joint appointments. We digress briefly to comment on this issue, before resuming our narrative.

In any modern university there are individuals or groups within departments other than Statistics who are either statisticians or probabilists themselves, or near enough to these to wish to have a close relationship with Statistics. This has long been the case in Berkeley, and Statistics has frequently reached out to people located in departments or units in the physical, engineering, biological, social, health and other sciences. Joint seminars, co-supervision and cross-listing of courses are some of the ways these interactions have been fostered. Sometimes individuals seek a more formal relationship, one permitting them to be the primary advisor of Statistics graduate students or to teach courses listed wholly within Statistics. In Berkeley there are two ways of responding to an interest of this kind: a joint appointment to two departments, usually organized when the person first arrives at Berkeley, or a zero-percent appointment in a second department, typically arranged for a person already at Berkeley.

We have many examples of both kinds of arrangements, going back to the mid-1950s, when the agricultural and resource economist George Kuznets (1909-1986) was the first zero-percent appointment in Statistics. Others from the past include Roy Radner, a statistician within the Department of Economics, now at New York University, Jacob Feldman, an ergodic theorist from the Department of Mathematics, Richard Barlow from Industrial Engineering and Operations Research, a Bayesian statistician interested in reliability, and Paul Holland, who was a statistician within our School of Education before returning to the Educational Testing Service. Currently we have eight zero-percent appointments: Kenneth Wachter from Demography, Leo Goodman from Sociology. Nicholas Jewell, Mark van der Laan and Sandrine Dudoit from Biostatistics, Alistair Sinclair from Computer Science (CS), Bernd Sturmfels from Mathematics, and Jasjeet Sekhon from Political Science. Albert Bowker (1919-2008), founder of the Stanford Statistics Department, became Berkeley Chancellor in 1971 with a 0% appointment in Statistics. In 1980 he moved office to the Statistics Department itself and interacted steadily with students and faculty alike. More generally, the participation of 0% faculty in our department's activities has greatly enriched the experience of our graduate students, as well as fostering closer collaborations among faculty. Conversely, several of our current faculty have 0% appointments in other departments, including Mathematics, CS and EE.

More challenging administratively, but potentially more effective at achieving the same aims, are formal joint appointments, usually 50:50. Our first such was Thomasian, who did his PhD with Blackwell on information theory, and later held a joint appointment between Statistics and EE.  In the mid-1990s Breiman urged the department to recognize the explosion of a kind of theoretical and applied statistics, growing within Computer Science departments, which contrasted dramatically with the view of the subject held in (most) Statistics departments. He argued strongly that we should narrow the gap between these two approaches by making suitable full or joint appointments of people with this background, usually termed machine learning. After considerable discussion between Statistics and CS and quite an extensive search, Michael Jordan was appointed in 1998 to do just this, and his arrival on the scene was quickly perceived to be a success. There are now four members of department with joint appointments of this kind: Jordan (joint with CS), with interests in machine learning, computational biology, Bayesian nonparametrics, graphical models, and natural language processing, Peter Bartlett (CS), with interests in artificial intelligence, machine learning, control and robotics, Martin Wainwright (EE), signal processing, coding, graphical models, optimization, sparsity in high-dimensional statistical inference, and nonparametric statistics, and Yun Song (CS), computational biology and population genetics. We also have two faculty joint with Mathematics, Michael Klass and Steven Evans, while Rasmus Nielsen is joint with Integrative Biology, with interests in evolutionary theory and genetics.

Returning to developments initiated around the year 2000, it is interesting to note that between 2003 and 2012, we made twelve new appointments including, two joint with CS, one joint with Electrical Engineering, three in applied and two in theoretical statistics, three in probability, and one joint with Integrative Biology. In the same period, several 0% appointments and adjunct appointments were made. Developments over this decade overseen by three different chairs seem to have followed the 2000 vision for "departmental renewal" surprisingly closely.

*Today*
The Department of Statistics at UC Berkeley currently has 43 faculty, comprising 13 with 100% Statistics appointments, 7 joint appointments spread across Mathematics (2), Electrical Engineering (1), Computer Science (3) and Integrative Biology (1), 3 with 100% teaching appointments (2 Senior Lecturers, 1 Continued Lecturer), 4 adjunct appointees in probability, statistics and finance, 8 zero percent appointments from within Biostatistics (3), Demography (1), Sociology (1), Computer Science (1), Mathematics (1) or Political Science (1), and 8 emeriti, 3 of whom are active within the Graduate School. In round numbers, we teach 2,000 students from other departments each semester, we have 250 students taking a BA majoring in Statistics, 10 students in our MA program, and 50 in our PhD program[21]. Furthermore, several of our PhD students specialize in either Communication, Computation and Statistics or Computational and Genomic Biology, these being two of the campus' cross-disciplinary PhD Designated Emphases. We are engaged in basic research in probability theory and mathematical statistics, including machine learning, and applied research in a wide range of areas noted earlier in this chapter.

We greatly value our interdisciplinary links and one of our goals is to maintain and extend these initiatives. Our graduate students are increasingly trained in a cross-disciplinary manner through collaborations with domain scientists and participation in the Designated Emphases mentioned above, and we aim to expand these efforts. We are also expanding our Masters program and updating its curriculum, and seeking to improve our interactions with business, industry and government, in an effort to meet the huge and growing demand for data analysis skills in these areas.

**Notes.**
There are published conversations with, volumes of papers honouring and obituaries of many former and current Berkeley Statistics faculty. We refer to *Statistical Science* for the first, and http://www.stat.berkeley.edu/people/memorials/ for the last mentioned. The remainder are readily accessible, e.g. through PROJECT euclid or JSTOR. We give birth dates only for deceased individuals at the first appearance of the name, and in those cases include the date of death. Moore (2007) is a further valuable reference.

1. Letter from Neyman to Deming (written c1937), quoted in Reid (1982), p.151.
2. Letter from Neyman to Provost Deutsch (written c1944), quoted in Reid (1982), p.197.
3. Neyman (1949), p.21.
4. Lehmann (2008), p.27.
5. Reid (1982, pp213 *et seq*), Lehmann (1996, pp142-143), Moore (2007, chapter 11), Lehmann (2008, p.91).
6. Evans had been keen to recruit a leading figure in Statistics to his department. He brought RA Fisher to Berkeley in September 1936 as Hitchcock Lecturer, but the visit was not a success, see Moore (2007, p.73). Later Raymond T Birge (1887-1980), chair of the Berkeley physics department recommended Neyman on the basis of the high opinion his collaborator W. Edwards Deming (1900-1993) had of him. Deming had been instrumental in organizing Neyman's 1938 *Lectures and Conferences*. He wrote a major paper with Birge, see the references.
7. Reid (1982, p.247), Lehmann (1996, p.143), Lehmann (2008, p.97).
8. Moore (2007, p.162), based on Reid (1982, pp.243-250).
9. Lehmann (2008), p.44.
10. Lehmann (2008), p.114.
11. Freedman et al (1978), p.xv.
12. Yang (1999), p.223.
13. Neyman (1949), p.27
14. Kendall *et al* (1982).
15. Neyman (1949), p.48.
16. Lehmann (2008), p.72.
17. Lehmann (2008), p.34
18. Neyman (1945), p.83.
19. Olshen (2001), p.191.
20. The approximately 500 Berkeley Statistics PhD graduates are all listed at http://www.stat.berkeley.edu/alumni/

# Appendix

*1. Books by Neyman and the founders*. For brevity we give first editions only. Many of these books have multiple editions, and translations into other languages. Also, we omit edited volumes, such as the *Proceedings of the Berkeley Symposia on Probability and Mathematical Statistics,* which are all readily available in PROJECT euclid.

J Neyman and E S Pearson. (1967). *Joint statistical papers of J. Neyman and E. S. Pearson.* University of California Press, iv+299
Henry Scheffé. (1959). *The analysis of variance.* John Wiley & Sons Inc., New York, xvi+477.
Charles Stein. (1986). *Approximate computation of expectations.* Institute of Mathematical Statistics, Hayward, CA, iv+164.

*Selected books by non-founders.* The books below are representative of a much larger number written by faculty in Berkeley Statistics. Again, we restrict to first editions only, and omit reference to translations, conference proceedings or collections of other kinds.

David J. Aldous. (1985). *Exchangeability and related topics.* École d'Été de probabilites de Saint-Flour, XIII, 1-198, Lecture Notes in Math., 1117. Springer, Berlin, 1-198.
Martin Anthony and Peter Bartlett (1999) *Neural Network Learning: Theoretical foundations.* Cambridge University Press. xiv+389 pp.
Peter J Bickel and Kjell A Doksum. (1976). *Mathematical statistics. Basic ideas and selected topics*, Holden-Day Series in Probability and Statistics Holden-Day Inc., San Francisco, Calif., xiv+493 pp.
Leo Breiman, Jerome H. Friedman, Richard A. Olshen, and Charles J. Stone. (1983) *Classification and Regression Trees*. Wadsworth Advanced Books and Software, Belmont, CA.
David R Brillinger. (1975). *Time series. Data analysis and theory*, International Series in Decision Processes Holt, Rinehart and Winston, Inc., New York, xii+500.
Ching-Shui Cheng (2013) *Theory of Factorial Design.* Chapman and Hall.
Lester E. Dubins and L. Jimmy Savage (1965) *How to Gamble if You Must: Inequalities for stochastic processes.* McGraw-Hill Book Co., New York-Toronto-London-Sydney. xiv+249 pp.
Steven N Evans. (2008). *Probability and real trees.* Lectures from the 35th Summer School on Probability Theory held in Saint-Flour, July 6-23, 2005 Springer, Berlin, xii+193.
David A Freedman. (2005). *Statistical models: theory and practice.* Cambridge University Press, Cambridge. x+414 pp.
David A Freedman. (2010). *Statistical models and causal inference.* A dialogue with the social sciences, Edited by David Collier, Jasjeet S. Sekhon and Philip B. Stark Cambridge University Press, Cambridge, xvi+399 pp.
David Freedman and Robert Pisani and Roger Purves. (1978). *Statistics.* W.W. Norton & Co.
Deborah Nolan and Terry Speed (2000) *Stat Labs: Mathematical Statistics through Applications*, Springer. xviii+ 300 pp.
Jim Pitman. (2006). *Combinatorial stochastic processes.* Lectures from the 32nd Summer School on Probability Theory held in Saint-Flour, July 7--24, 2002. Springer-Verlag, Berlin, x+256.
Jim Pitman. (1993). *Probability.* Springer, xi+559 pp.
John A. Rice. (2006). *Mathematical statistics and data analysis.* Cengage Learning, 666 pp.
P. B.Stark (1997). SticiGui: Statistics Tools for Internet and Classroom Instruction with a Graphical User Interface. http://statistics.berkeley.edu/~stark/SticiGui
Charles Stone (1995) *A Course in Probability and Statistics.* Duxbury, 838 pp.
Aram J. Thomasian. (1969). *The structure of probability theory with applications.* McGraw-Hill, 746 pp.
Martin J. Wainwright and Michael I. Jordan (2008). Graphical Models, Exponential Families, and Variational Inference. *Foundations and Trends in Machine Learning.* Vol. 1: No 1-2, pp 1-305.

*Photographs*

Stat. Lab. Univ. of Calif. Summer Session 1947.
Names: (left to right) 1. E. L. Crow 2. E. S. Keeping 3. V. Lenzen 4. D. G. Chapman 5. C. Stein 6. E. Crow 7. E. Fix 8. S. W. Nash 9. E. L. Scott 10. D. Cruden Lowry 11. J. Gurland 12. I. Elvebach 13. E. L. Lehmann 14. T. A. Jeeves 15. E. Seiden 16. E. A. Fay 17. B. Epstein 18. T. Birge 19. I. Blumen 20. Unknown 21. B. M. Bennett 22. J. L. Hodges 23. T. Hodges 24. H. Cramer 25. G. R. Seth 26. J. Neyman 27. H. Hotelling 28. Z. Szatrowski.

For more, go to: http://www.stat.berkeley.edu/photos

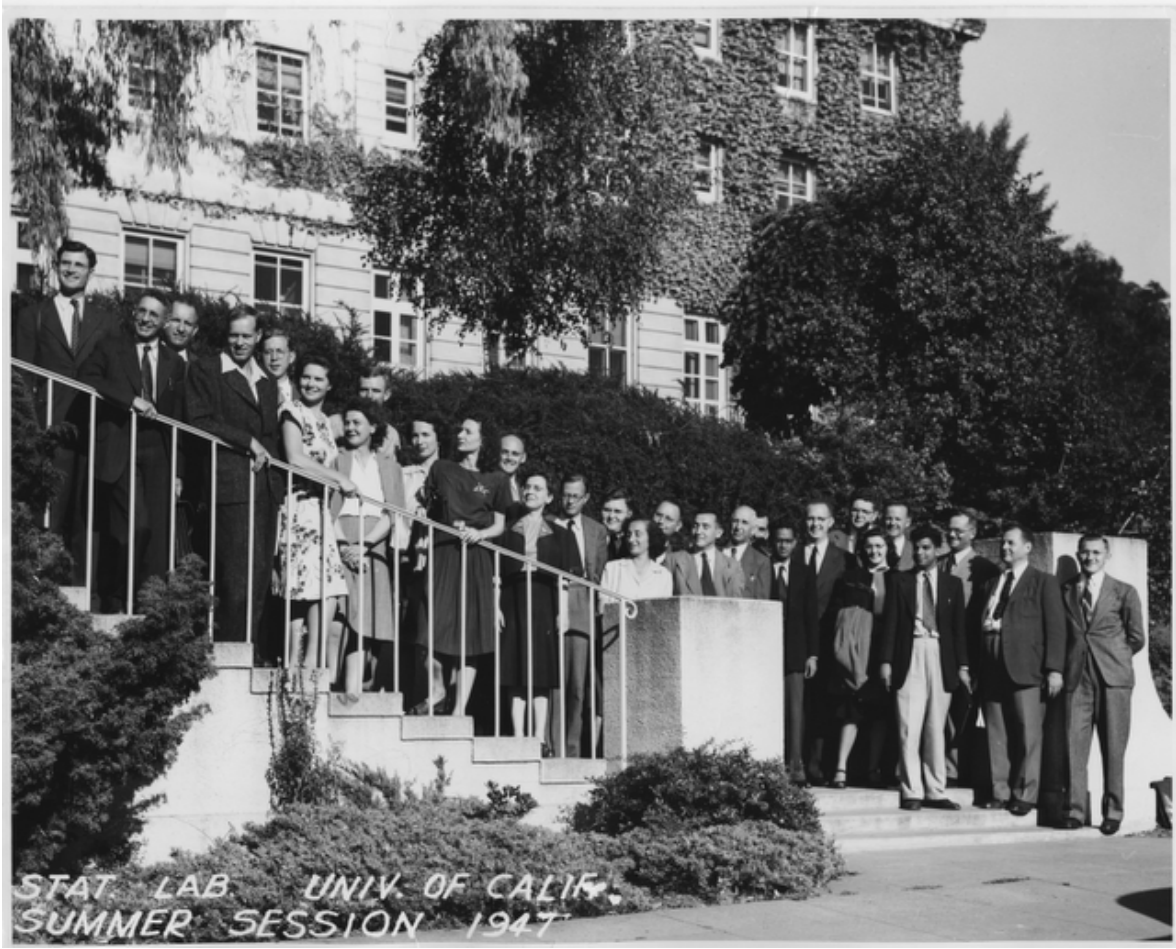

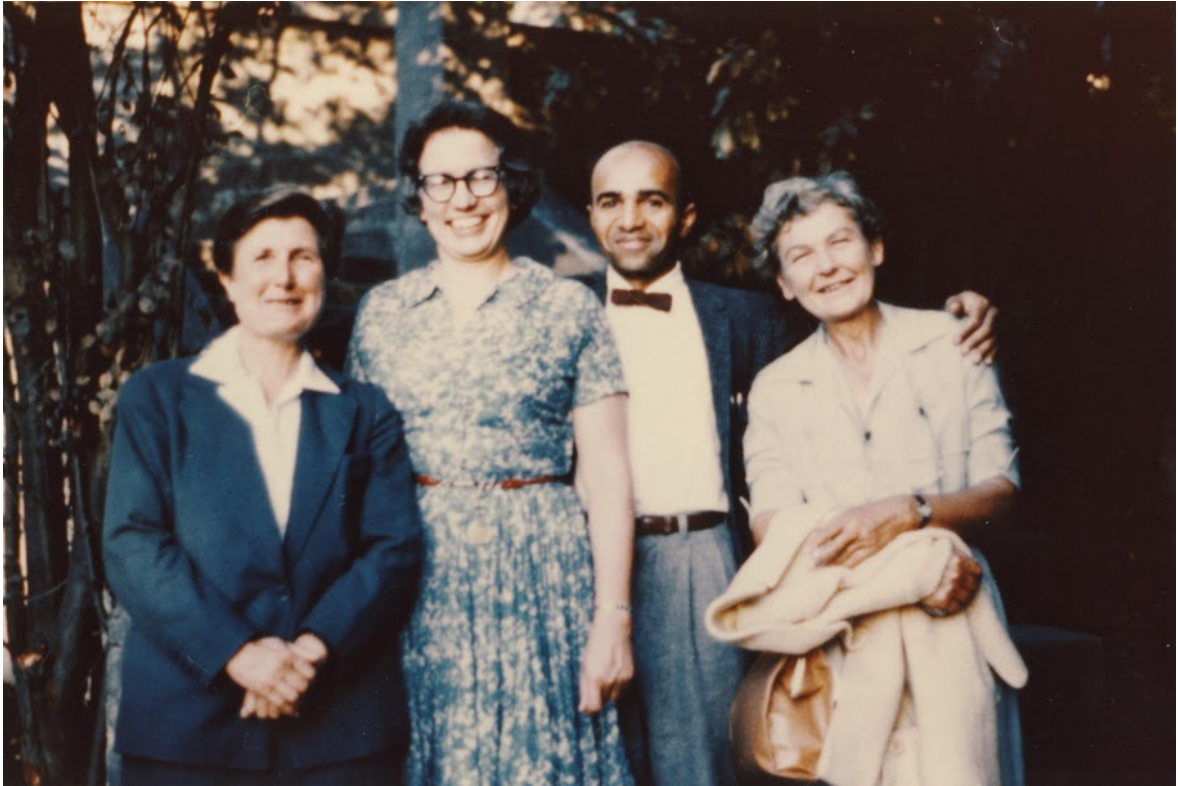

FN David, EL Scott, D Blackwell, E. Fix.

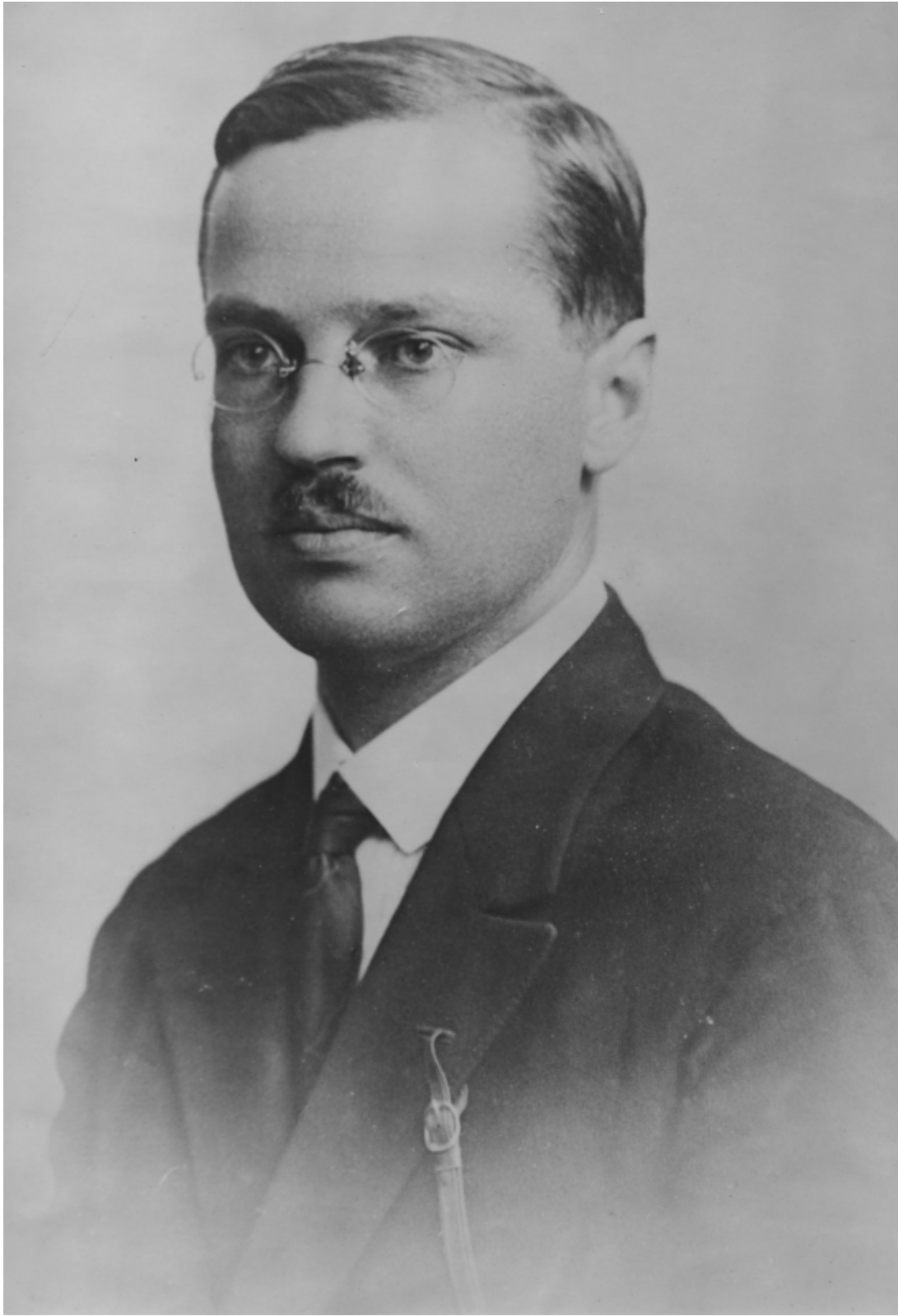

J Neyman

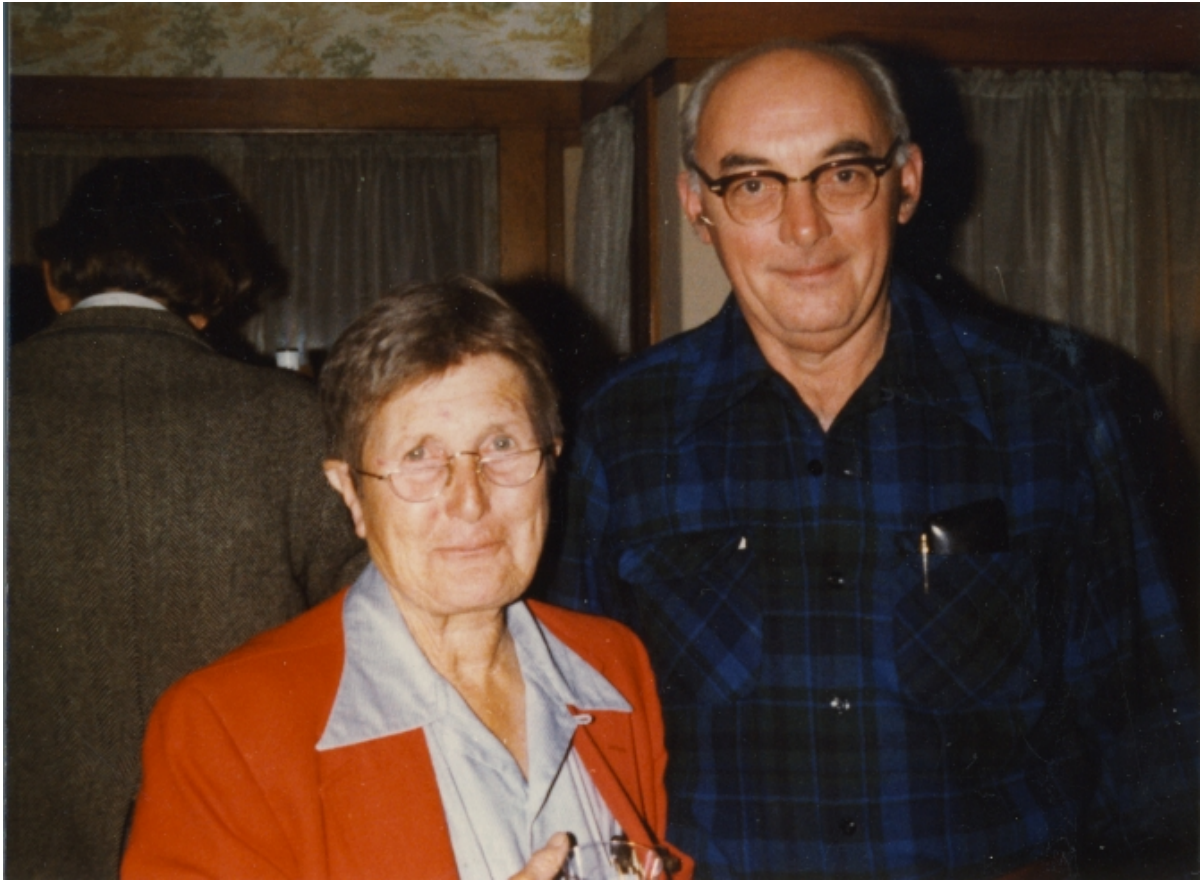
FN David and L LeCam

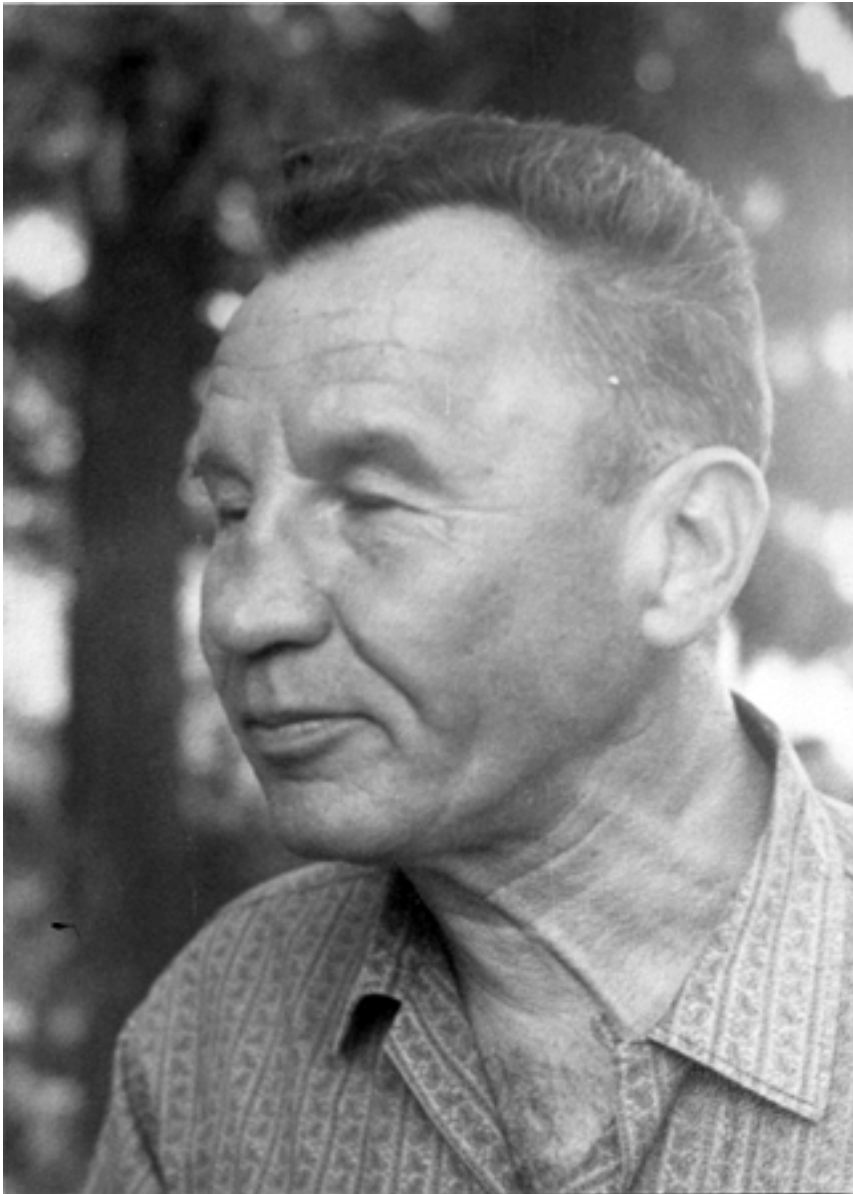
H Scheffé

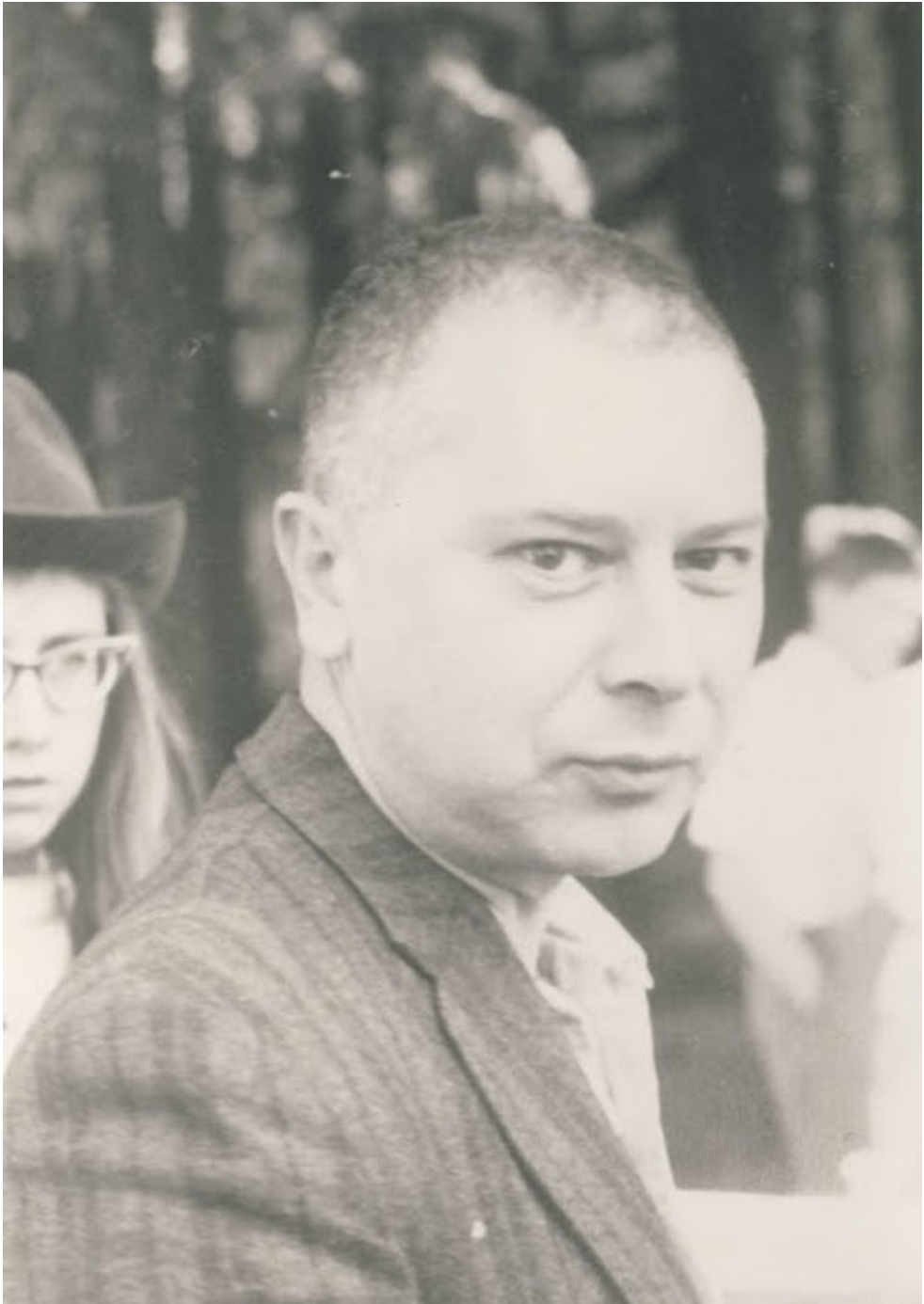
E Barankin

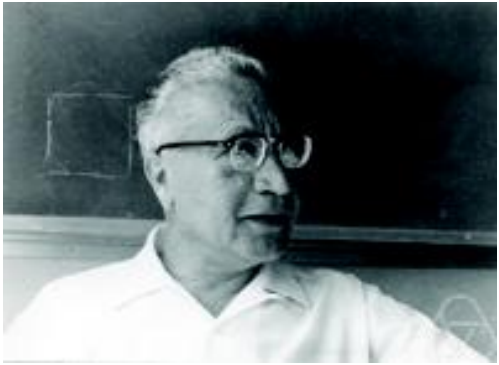
M Loève